\documentclass[conference]{IEEEtran}
\IEEEoverridecommandlockouts
\usepackage{cite}
\usepackage{amsmath,amssymb,amsfonts}
\usepackage{algorithmic}
\usepackage{graphicx}
\usepackage{physics}
\usepackage{textcomp}
\usepackage{xcolor}
\usepackage{hyperref}
\def\BibTeX{{\rm B\kern-.05em{\sc i\kern-.025em b}\kern-.08em
    T\kern-.1667em\lower.7ex\hbox{E}\kern-.125emX}}
\begin{document}

\title{Control Analysis of Packet Transmission Algorithms: Study on Fairness and Stability}

\author{\IEEEauthorblockN{Lokesh Bommisetty}
\IEEEauthorblockA{\textit{Department of Electrical Engineering} \\
\textit{Indian Institute of Technology Madras}\\
ee18d412@smail.iitm.ac.in}}

\maketitle

\begin{abstract}
This document is a study of fairness, feedback and stability notions of different packet transmission algorithms. We start the discussion with defining two scalable control algorithms namely primal and dual algorithm. We discuss the dual algorithm model and then understand the fair dual algorithm. Further, we discuss different notions of fairness under fair dual algorithm those correspond to TCP and RCP congestion control protocols. Feedback parameters are analyzed in each of these fairness algorithms and thus their stability is studied.   
\end{abstract}

\section{Introduction}
Understanding the dynamics of congestion in the network has been a prime interest if research. The primary objective of any congestion control algorithm is to utilize the bandwidth efficiently and fairness in resource allocation to end systems.  To have a better control over the congestion in the network, the congestion control algorithms are coupled with the feedback from the network to facilitate the transmission rates of end systems adapt themselves according to congestion in the network. These feedback mechanisms are referred to as active queue management (AQM) \cite{b1}. 

To analyze these congestion algorithms, we convert them into delay differential equations or fluid models. \cite{b2} shows that the current congestion avoidance TCP, when coupled with commonly used AQM, loses local stability with increase in delays. In the process of finding an algorithm whose stability properties are invariant to delay, network topology, number of users and capacity led to the proposition of two scalable control algorithms which can be classified into primal and dual algorithms \cite{b3}\cite{b4}. This averaging can be done at both end systems and resources. Such systems correspond to the primal-dual framework. \cite{b7} provides overview of these primal, dual and primal-dual algorithms and their local stability properties.
 
In the primal congestion control algorithm, the congestion controllers adjust their transmission rates after averaging the congestion feedback signals from the network while in the dual version of algorithm this averaging is performed at the resources \cite{b1}. At this point, we briefly recall the Transmission Control Protocol (TCP) and Rate Control Protocol (RCP). The TCP framework has embedded in it a mechanism for detecting congestion and then it adjusts the transmission window sizes at the end systems. TCP follows a conservative increase of window size and aggressive decrease of window size on detection of congestion. While in sharp contrast, RCP aims by fast flow completion times by communicating explicit rate feedback between end users and resources \cite{b5}\cite{b6}. Dual algorithm broadly correspond with more explicit congestion protocols \cite{b8}.

In this paper, we discuss three fair dual algorithms. These are fair dual algorithms with proportional fairness, TCP fairness and max-min fairness which discuss the different strategies of fair resource allocation. For this we first model the dual algorithms where we model the system with non-linear delay equations and use bifurcation theoretic tools to get an insight into the dynamics of these algorithms. From this model of dual algorithm, we introduce the notion of proportional fairness, TCP fairness and max-min fairness under different choices of parameters of the model. We then analyze the RCP model under the same parameters as considered in dual algorithm and show that RCP can exhibit max-min fairness.

Linearized analysis offers conditions for local stability, but congestion controllers may become locally unstable under the bifurcation. This does not imply that the non-linear system would have unbounded oscillations. So we take non-linear delay equations by allowing linear, quadratic and cubic terms and then apply bifurcation theory to analyze the stability based on the amplitude of oscillations.

The rest of the paper is organized as follows. In Section II, we give a brief modelling of the dual algorithm to introduce the $\alpha$-fairness. In Section III we discuss proportional, TCP and max-min fairness. In Section IV we study different feedbacks we encountered in the modelling of networks and study how they are helpful. In Section V we analyze the stability of the system under different fairness algorithms and finally, we conclude the paper in Section VI.  
\section{Dual Algorithms}
Dual algorithms are a class of congestion control algorithms where resource determines the congestion measure hereafter called as price, by averaging the congestion feedback before communicating it to the end systems as discussed in Section I. This feedback would prompt the end systems to raise a response for their demand for a share of resource which is beneficial for networks with shorter round trip times. 

Fair allocation of resources to all the end systems is also a key concern. Some of the notions of fairness are i) proportional fairness, ii) TCP fairness iii) max-min fairness. Proportional fairness is attractive from economic point of view. TCP fairness is offered by the TCP protocol. Max-min fairness is about maximizing the minimum share of resources to the end systems. Max-min fairness is exhibited by RCP. We will discuss these notions of fairness in detail later.

In this paper we consider the users face a forward delay (end system to the link) of $\tau^f$ and a backward delay (link to end system) of $\tau^b$. The round trip time propagation delay assuming no other delays is the sum of forward and backward delay, given as $\tau = \tau^f + \tau^b$. Then the system equation is written as:\\
\begin{equation} \label{1}
    \dv{}{t}x(t) = \kappa f(x(t),x(t-\tau))
\end{equation}
Considering the dynamical representation of dual algorithm \eqref{1} is modified as 
\begin{equation} \label{2}
    \dv{}{t}p(t) = \kappa p(t)^m (x(t-\tau)-CI_{[p(t)>0})
\end{equation}
where \begin{equation}\label{3}
    x(t) = \mathcal{D}(p(t))
\end{equation} 
with $\mathcal{D}(p),p\geq0$, a non negative continuous, strictly decreasing demand function. $C$ is the $capacity$, $p$ is the $price$ at the link and $\kappa$ is the gain of the system. Any system of the form in \eqref{2} is referred to as a dual algorithm. In \eqref{2} if $m=0$ then the system is said to be delay dual and if $m=1$, it is said to be fair dual. The demand function of fair dual corresponds to the demand function 
\begin{equation}
    \label{4}
    \mathcal{D}(p) = \left(\frac{w}{p}\right)^{1/\alpha}
\end{equation}
which achieves $\alpha$-fair allocations \cite{b7}. $w$ is the $willingness to pay$ parameter of the user. With these results the model of fair dual algorithm reduces to 
\begin{equation}
    \label{5}
    \dv{}{t}p(t) = \kappa p(t) (x(t-\tau)-C)
\end{equation}
where $x(t-\tau) = (w/p(t-\tau))^{1/\alpha}$ according to \eqref{3},\eqref{4}. 
\section{Fairness}
The major communication network design challenge is sharing the available resource (bandwidth) between the competing users of the network while ensuring the maximum utilization of the network. As discussed in Section II the fair dual algorithm is represented by \eqref{5} for $\alpha$ fair allocations. This parameter $\alpha$ can take values in [0,$\infty$). $\alpha$ = 0 corresponds to maximum throughput of the network. We will discuss the different notions of fairness for different choices of $\alpha$ as follows.
\subsection{Proportional fairness}
Proportional fairness ensures that the resources are allocated to users proportionally. It achieves a good trade-off between efficiency and fairness. \cite{b9} concludes that the proportional fairness solution simultaneously achieves higher system throughput, better fairness and lower outage probability with respect to the default solution given by today's 802.11 commercial products \cite{b8}. \cite{b10} describes a model for elastic traffic in which a user chooses the charge per unit time that the user is willing to pay; thereafter the
user's rate is determined by the network according to a
proportional fairness criterion applied to the rate per unit
charge. 

With the parameter choice of $\alpha = 1$ and $w = 1$ in \eqref{4} we get a proportionally fair resource allocation, which yields 

\begin{center}
    $x = \frac{1}{p}$
\end{center}

Consider a  network has routes $r:$ $r \in R$ where $R$ is the collection of all possible routes. A vector of transmission rates $\Bar{x} = (x_r, r \in R)$ is said to be proportionally fair if under the conditions $x\geq 0$ and $Ax\leq C$, $\sum_{r \in R} \mathcal{U}_r(x_r) $ is maximum. Here $A$ is the incidence matrix and $\mathcal{U}_r$ is is the utility of the $r^{th}$ user as defined in \cite{b8}. This is verified as for any other vector $x^*$ the aggregate proportional changes in negative or zero:
\begin{center}
    $\sum_{r\in R} \frac{x^*_r - x_r}{x_r} \leq 0$
\end{center}

\subsection{TCP fairness}
Most of the applications in the world today use TCP to transmit data over the network. TCP is a congestion control protocol which means it adapts the transmission rate depending on the congestion in the network. TCP adapts a conservative increase and multiplicative decrease pattern of congestion window. When the network is free and no congestion is detected by TCP, the transmitting window size is increased by one for every round trip time and if it detects congestion, window size is reduced to half. In this process, fairness would refer to modelling of window sizes of different users according to congestion.For example in a network of 10 flows, if the congestion in the network is caused by a single flow then the fairness part comes into play in deciding how to back-off the transmitting rate. It can be done by backing off all the flow transmitting rates or by backing off the single flow rate that causes congestion. But the later requires the knowledge of the transmitting rates of all the flows. If the transmitting rate of each user is also made available in the network, that may cause congestion in the network by itself.

In \eqref{4}, if we choose $\alpha = 2$, $w = 1/\tau^2$ then the corresponding fair dual algorithm is said to have TCP fairness, which yields the transmitting rate to be
\begin{center}
    $x = \frac{1}{\tau\sqrt{p}}$
\end{center}
which is same as the equilibrium condition of the TCP fluid model. So we can say that TCP flows exhibit TCP fairness.
\subsection{Max-min fairness}
In a network, max-min fairness corresponds to the fairness algorithm that maximizes the minimum share of resource allocation to every user of the network. In detail consider a network consists of interconnection of a set of routes, $S$, with a set of links, $J$ as described in the RCP model in \cite{b11}. Every packet transmitting through a link carries in its header, an explicit rate feedback variable. Initially this variable is set to the desired flow rate for route r. As the packet passes through each link, it will be replaced by the maximum bandwidth $R_l(t)$that link can allocate to the particular route. If multiple routes pass through the same link, then the link capacity is thus distributed among all the routes i.e, $\sum_{r_l} R_{lr} = C_l$ where $R_{lr}$ is the bandwidth share of route $r$ passing through link $l$, $C_l$ is the capacity of link $l$ and $r_l$ is the collection of routes passing through link $l$. 

For a particular route, once a packet reaches the destination, explicit rate feedback is sent to the user so that the transmission rate can be modelled according to the available resource to it. For a particular link, as it sees multiple packets of different routes carry different demand rates, the allocation share keeps on changing which is modelled in  \cite{11}. After multiple iterations of packet flows through the network the transmission rate of a flow converges to the equation 
\begin{equation} \label{6}
    x_r(t) = \left(\sum_{j\in J} R_j(t-T_{jr})^{-\alpha}\right)^{-1/\alpha}
\end{equation}
where $T_{jr}$ is the return delay from link $j$ to to source of route $r$.  \eqref{6} depicts the $L_\alpha$ norm expression of a sequence $\frac{1}{R_j(t-T_{jr})} $ in $j$. As $\alpha\rightarrow\infty$, the $L_\infty$ norm takes the maximum value of the sequence \cite{b13}. Which means that the transmission rate of a user converges as

\begin{equation*}
x_r = \underset{j \in J} max \left(\frac{1}{R_j(t-T_{jr})}\right)
\end{equation*}
\begin{equation}
    x_r = \underset{j \in J}min (R_j(t-T_{jr}))
\end{equation}
which is same as choosing the parameter $\alpha \rightarrow \infty$ in \eqref{4}. 
\section{Feedback}
Fluid model of RCP \cite{b11} is given as ,
\begin{equation}
\dfrac{dR_l(t)}{dt}=R_l(t) \left(\dfrac{\alpha_l}{d_lC_l}(C_l-y_l(t))-\dfrac{\beta_lq_l(t)}{d_l^2C_l}\right)_{R_l(t)}^+ \label{eq2.4}
\end{equation}
where $R_l$ is the flow rate associated with link \textit{l}, $\alpha_l$,$\beta_l$ are positive constants, $C_l$ is the capacity at link \textit{l}, $d_l$ is the average RTT of flow passing through link \textit{l}, \textit{$q_l(t)$} is the queue size at time t, and $y_l(t)$ is the aggregate flow rate of all the link. Notation $a = (b)^+_c$ means that $a=0$ if $b<0$ and $c\leq0$, otherwise $a = b$.

The fluid model of RCP \eqref{eq2.4} contain two feedback terms. First term ($C_l$ - $y_l(t)$) is known as rate mismatch which is the difference between the link capacity $C_l$ and aggregate arrival rate $y_l(t)$ and second term is based on queue size $\left(\dfrac{\beta_lq_l(t)}{d_l^2C_l}\right)$. Apart from these two, other feedbacks like packet loss, rate, etc. are also available. The systems with and without feedback based on queue size give rise to different nonlinear equations, so these feedback factor are required to develop sufficient condition for local stability. Packet loss feedback tell the sender to re transmit the packets. Rate feedback is used to avoid congestion in the network.

The feedback is necessary to ensure that the network is stable and hence we can derive the necessary conditions and sufficient conditions of different parameters of the network for stability. Feedback is required for an end system to decide upon its transmitting rate. Feedback is also necessary to model algorithm for resource sharing. For example, TCP fairness algorithm uses packet loss as a feedback. If any congestion is detected in the network, TCP will reduces its window size to half to avoid congestion in the network and if no congestion is detected in the network then window sized will be increases in every RTT interval. In compound TCP congestion in the network as well as delay feedback is available to model window size. In RCP, demand rate of user is taken as input for modelling the bandwidth share allocated to that particular user which is known as max-min fairness. In case of proportional fairness bandwidth is distributed according to amount pay by user.

\section{Stability Analysis}
We will discuss the stability analysis of different fairness algorithms in this Section. We will derive the notion of stability using linear system analysis and bifurcation analysis for TCP and RCP models.
\subsection{Stability analysis of TCP}
The fluid model for TCP is derived and formulated in \cite{b12} as 
\begin{equation}
\dfrac{dw(t)}{dt}=\dfrac{1}{RTT}-\dfrac{w(t)}{2} \left[x(t-RTT)p(t-RTT)\right] \label{9}
\end{equation}
where \textit{w(t)} is the average window size of \textit{N} flows at time \textit{t}, measured in packets, \textit{x(t)} average rate at which packets are transmitted at time \textit{t}. Average transmitting rate and average window size ar related as $x(t) = w(t)/RTT$, \textit{p(t)} packet loss probability at time \textit{t}.

The stability of system is calculated by finding the equilibrium point of \eqref{9}
and by finding the packet loss probability at the equilibrium point as discussed in \cite{b12}. From conclusion from \cite{b12}, if traffic intensity ($\rho$) is less then 1, then the system is stable, for small buffer, for all window sizes where $\rho = x(t)/C$. But if we increase the buffer size, choice of window size is get limited, but utilization gets increased \cite{b14}.

Relationship between stability and desynchronization is shown in \cite{b14}, which says that for a large buffer size, we can see oscillations in $\rho$ around $\rho = 1$. When $\rho\geq 1$ the queue size bounces around full and packet loss occurs. When $\rho < 1$,the queue size bounces around empty. So change in queue size is very large even with small change in $\rho$. When $\rho ~ 1$, the is queue is full and all the flows experience drop at nearly the same time and so all the flows back-off and hence get synchronized. Hence we can say that large buffers promote synchronization as pointed in \cite{b14}. For a small buffer, fluctuation is small in $\rho$ and queue size has large and fast variation. Fluctuations or oscillation in $\rho$ correspond the amplitude of limit cycles. TCP cannot control the fluctuations in queue size, but its distribution can be controlled. 

\subsection{Stability analysis of a RCP}
Fluid model of RCP is explained in \cite{b11} as
\begin{equation}
\dfrac{dR_l(t)}{dt}=R_l(t) \left(\dfrac{\alpha_l}{d_lC_l}(C_l-y_l(t))-\dfrac{\beta_lq_l(t)}{d_l^2C_l}\right)_{R_l(t)}^+ \label{eq3.4}
\end{equation}
where $R_l$ is the flow rate associated with link \textit{l}, $\alpha_l$,$\beta_l$ are positive constants, $C_l$ is the capacity at link \textit{l}, $d_l$ is the average RTT of flow passing through link \textit{l}, \textit{$q_l(t)$} is the queue size at time t, and $y_l(t)$ is given as 
\begin{equation}
y_l(t) = \sum_{r:l\epsilon r} x_r(t-\tau_{rl}) \label{eq3.5}
\end{equation}
where $\tau_{rl}$ denote the propagation delay from the origin of origin r to link l. The notation a = $(b)^+_c$ mean that a=0 if b less then 0 and c $\leq$ 0, otherwise a = b.
For small buffer $q_l(t)$ is given as 
\begin{equation}
q_l(t) = p_l(y_l(t) \label{eq3.6}
\end{equation}
It is proven in \cite{b11}, for delay free model of RCP, at equilibrium, is globally stable, but author also show that equilibrium point of \eqref{eq3.4} - \eqref{eq3.6} cannot be unique, or even isolated. If $\alpha_l$ and $\beta_l$ in \eqref{eq3.4} is small then algorithm evolves slower then propagation delay, therefore \cite{b11} conclude that with sufficiently small $\alpha_l$ RCP is globally stable.

Local stability can be found by local Bifurcation analysis \cite{b11}\cite{b12}. For single link single delay model, and taking $\beta_l$ = 0 for sake of simplicity, RCP fluid model is rewritten as 
\begin{equation}
\dfrac{dR(t)}{dt} = \eta R(t)\left(\dfrac{\alpha}{C_{\tau}}(C-y(t))\right) \label{eq3.7}
\end{equation}
And necessary and sufficient condition \cite{b11} for local stability is 
\begin{equation}
\eta \leq  \dfrac{\pi}{2}
\end{equation}

As the notions of proportional, TCP and max-min fairness is derived from the generalized dual algorithm model, we will compare the local stability in terms of the parameter $\alpha$ as discussed at the end of Section II which corresponds to amplitude of limit cycles. \cite{b1} derives that the leading oscillation coefficient of a fair dual algorithm is proportional to $\alpha$. We understood that proportional fairness corresponds to $\alpha = 1$, TCP fairness correspond to $\alpha = 2$. So, the TCP fair algorithm have a limit cycle with twice the amplitude of its proportionally fair counterpart \cite{b1} while max-min fairness will have oscillation coefficient proportional to the average allocated rate as discussed in \cite{b11}.

\section{Conclusion}

In this paper, we started our discussion with the introduction of two scalable control algorithms primal and dual algorithms. We then modelled the dual algorithm and defined the origin of fair dual algorithms. Saying that the fair dual algorithm ensures $\alpha$ fair allocations, we then explained different notions of fairness for three different choices of $\alpha$. We discussed the fairness exhibited by proportional, TCP and max-min fairness algorithms. We then discussed the feedbacks available in each of the algorithms and concluded with the stability analysis of each fairness algorithm corresponding to its feedback available in their network.

\end{document}